# Observations of Loki Patera by Juno during close flybys

A. Mura[1], and the JIRAM-Io team

1 Istituto Nazionale di Astrofisica – Istituto di Astrofisica e Planetologia Spaziali, Rome, Italy

**Abstract.** We used data from the Juno spacecraft to investigate both the spatial and temporal properties of Loki Patera on Io, acquired in two infrared bands between December 2022 and April 2024, at spatial resolutions ranging from 400 m to 15 km. Loki shows a thermal structure unlike other active lava lakes previously reported, with some brightening near the lake's perimeter but lacking the continuous "hot ring" seen at other paterae. Modeling the slow rate of cooling suggests there is a significant volume of magma beneath the crust to provide the latent heat necessary to decelerate the cooling rate.

A thermal propagation that may represent the signature of a resurfacing wave, going from the southwest of the lake to the north, was observed with a velocity of ~2–3 km/day. Data collected in 2024 may indicate the onset of a new resurfacing wave originating from a point source, rather than the foundering of a linear section of the crust.

We also observed many small (~3 km wide), closely spaced (~ 10 km apart) islands that have persisted in the same locations for at least 45 years, since first being imaged by Voyager 1. The persistence of these islands challenges resurfacing models of Loki, as they have remained fixed—likely anchored to the lava lake floor—and have not noticeably changed in size, arguing against large-scale thermal erosion. The central island of Loki shows a few thermal structures associated with the fractures that cross the island, indicating that the fractures most likely contain molten lava.

**Key points:**

- Many small (~3 km wide), closely spaced islands have persisted for at least 45 years.
- A thermal signature that may be a resurfacing wave is observed; the velocity of such wave is constant at ~2-3 km/day.
- The cooling of the crust is slow and implies that a significant amount of hot lava exists beneath the crust, which has implications for the depth of the active lava lake.
- Considerable heat emission is found near the edges of the patera, but with a more complex pattern compared to hot rings observed in other lava lakes on Io.

- The central island is warmer than the terrain surrounding Loki Patera.
- The average total power emission from Loki Patera is estimated as 10 TW, in good agreement with previous estimates.

# 1 Introduction

Loki Patera is the largest volcanic depression on Jupiter's moon Io, making it a focal point for studying extra-terrestrial volcanism (Lopes and Williams, 2015). Observations by the Voyager 1 and Galileo spacecraft revealed its unique characteristics: it is a dynamic volcanic region that spans approximately 200 kilometers in diameter, covers an area of around 21,500 square kilometers, and has a long-lived large inner island. This enigmatic patera is where active lava prevails in the form of extensive lava flows across the floor of the depression, or perhaps as a vast lava lake (or sea), characterized by quasi-periodic brightening events (Rathbun et al., 2002; 2006; de Kleer and Rathbun, 2023). Being the largest and often the most luminous hotspot on Io, Loki Patera is the most thoroughly examined (Rathbun et al., 2002; Gregg and Lopes, 2004; Matson et al., 2006; de Kleer et al., 2017), though without high-resolution imaging, and the exact mechanisms driving its activity are still not fully understood. The level of activity at Loki Patera produces ~10% of the total thermal emission from Io (de Kleer and Rathbun, 2023).

It is unclear how much Loki Patera's behavior is similar to that of other, smaller hotspots, and what factors drive its activity compared to other lava lakes on Io. According to Rathbun et al. (2002), observations made between 1988 and 2000 show that Loki Patera experienced significant brightening roughly every 540 days, with each bright phase lasting around 230 days, followed by a recovery phase for the remainder of the period. Temperature maps derived from Galileo observations in 1999 and 2001 support the idea of a consistent resurfacing wave moving at 1-2 km per day, with a counterclockwise direction starting in the south, as observed by Voyager 1 and Galileo (e.g. Spencer et al., 2000; Howell and Lopes, 2007). However, adaptive optics data from Keck and Gemini telescopes (de Kleer et al. 2017) suggest a clockwise wave starting in the north, indicating the presence of two independent resurfacing waves (de Kleer et al., 2017). Despite some variability in each brightening cycle, all past observations align with the progression of the resurfacing wave and occasional activity at the southwestern margin of Loki Patera.

Rathbun et al. (2002) and Matson et al. (2006) proposed that this wave signifies the collapse of the solid crust of a silicate lava lake, with the periodicity reflecting the time required for the crust to grow thick enough to overturn due to a density imbalance. The lava lake model is further supported by analysis and modeling of near-infrared mapping spectrometer (NIMS) data by Lopes-Gautier et al. (2000), and Howell and Lopes (2007), and by ground-based observations by de Kleer and de Pater (2017), and by de Kleer et al. (2017). However, comparisons with terrestrial lava lakes (Gregg and Lopes, 2008) show notable differences between Io and the Earth and suggests instead a mechanism in which the lava forms a temporary lava lake before draining back.

Complicating matters, in 2000, Loki Patera's semiregular brightening pattern appeared to stop, shifting to a more constant but moderate level of activity (Rathbun and Spencer, 2006) until 2013 when it again returned to quasi-periodic eruptions, but with a shorter period (de Kleer and Rathbun, 2023).

Understanding the mechanisms behind these quasi-periodic eruptions provides critical insights into Io's geologic activity and offers broader implications for the longevity and magma recycling associated with terrestrial lava lakes (Harris et al., 1999; Harris, 2008). This paper delves into the physical and thermal properties of Loki Patera, analyzing its cyclic behavior, the role of magma dynamics, and the factors influencing its volcanic activity, as observed by Juno during some very close encounters of the spacecraft with Io.

## 2 Observations

In December 2023 and in February 2024, Juno made two close flybys of Io at a distance of ~3,500 km to 4,500 km, allowing remote sensing instruments to take images of the surface. JIRAM (Adriani et al., 2017) is an imager and a spectrometer in the 2 to 5 μm infrared wavelength range. Together with data from these close flybys, obtained in orbits 57 and 58, images from orbits 47, 51, 55, and 60 are used in this study to produce radiance maps of Loki Patera in both the L and M bands (see Appendix A1 for a description of the instrument and dataset) and to retrieve the surface temperature. Table A1 in Appendix A1 provides a list of the observations.

The five M-band radiance maps of Loki Patera taken from May 2023 to April 2024 (Figure 1) clearly show a continuous decrease in brightness temperature (see the additional color bar to the right), suggesting the recovery phase of one cycle of activity of the lake. Based on the measured temperatures and a model of magma cooling (see Appendix A2), we argue that this cycle started near December 2022 (orbit 47, Dec. 15, 2022), when the signal from Loki Patera in the M band was in fact very high and saturated. Due to the poor spatial resolution (Loki Patera was very close to the limb in that observation, and was just a few pixels across), the radiance map from orbit 47 is not shown here.

Figure 1 shows that Loki Patera is brightest in the north and dimmest in the southwest. Mura et al. (2024b) derived temperatures from M- and L-band radiance maps by using the L-over-M ratio, which is a function of the temperature. However, this method can be quite noisy if the radiance maps have high spatial resolution and are rich in features. In fact, even minor misalignment between the maps can create significant artifacts, which can affect the quantitative analysis we want to perform. However, in the case of the large areas covered by crust, a simple estimation of the brightness temperature (see below) is sufficient for our purposes. For these reasons we reduce the resolution of our dataset and analyze three specific zones that we will call zones 1, 2 and 3. Zone 1 is to the south, zone 2 to the north, and zone 3 to the east of Loki Patera (due to the lack of L-band data from orbit 57, we use the corresponding panel of Figure 1 to show the location of these zones).

We extract the median value of the radiance in each of these zones and derive the temperature by assuming a black body with emissivity ($\varepsilon$) as a function of the temperature (Ramsey et al., 2019). The result, temperatures as a function of time, is shown in Figure 2. Temperatures are correlated with the radiances observed in Figure 1 with the warmest areas in the north (zone 3) and the coolest in the south (zone 1).

In Figure 3 we show a comparison between the radiance map of Loki Patera (panel B), obtained by the JIRAM M-band imager on December 30, 2023, and the visible (VIS) counterpart (panel A), obtained by Voyager 1 in March 1979, over 40 years prior. While the coverage of the IR images is incomplete, the spatial resolution of this map (see Table 1 in Methods) is the highest one. The comparison reveals that there are many small spots, with very low M-band radiance, that align very well with bright spots in the VIS map. In Figure 4, we present three panels showing magnified regions of Loki Patera. For convenience, and since the three regions are located to the north, the south, and the east of Loki Patera, similarly to the zones in Figure 1, we name them with the same numbers.

In region 1, we identified ~10 spots that are coaligned well (dark in the IR, bright in the VIS); a similar analysis was conducted in region 2. In region 3, the spatial resolution of the images, combined with the large number of spots, makes comparison more challenging, and no co-aligned features were identified. Because of the very long time elapsed between the VIS and

IR observations, the possibility that these are lava rafts or crustal rafts is not compatible with the crustal overturn model since they should have moved due to the numerous overturning waves over approximately 45 years. Instead, we propose that these are islands anchored to the magma lake floor. The distance between these features, in the region where they are packed, is around 10 km or less, while their typical width is ~3 km; both values may be overestimated, due to the spatial resolution of JIRAM (though the size is also determined from the Voyager 1 visible image). There is insufficient information to determine the height of these features, but their longevity most likely constrains variability in the depth of the lava lake over the past 45 years.

## 3 Analysis

### 3.1 Morphology of Loki Patera in the IR.

Conrad et al. (2015) showed that detailed views of Loki Patera, captured through ground-based direct imaging, reveal emissions from two distinct areas within the patera's floor. According to their findings, the feature in the southwest corner likely represents a persistent high-temperature source, while the emission in the southeast is probably due to radiation from the nearby cooling crust.

However, the much higher-spatial resolution observations from JIRAM reveal a more complex morphology. The maps from May 2023 (panels A and F of Figure 1) show a maximum radiance in the northern part of the lake, which partially saturates the M-band channel. We interpret this as relatively young crust floating above underlying magma and cooling. In the maps from December 2023 and February 2024, which have very good spatial resolutions (orbits 57 and 58; panels C, D, I of Figure 1), the crust appears to have cooled, but there is a maximum radiant emission near the patera's wall to the south, like what was observed in NIMS data (Lopes et al., 2004). In the same maps, some regions of intensification of the brightness are visible close to the promontories that are at ~15°N and ~16°N to the east of the lake; these are more evident in Figure 3.

Additionally, two clear signatures of newly exposed lava were observed in April 2024 (panels E and L of Figure 1): one to the west near the patera's wall and one to the east close to the island's shore. In the earlier observation from February 2004, a minor lava feature was visible to the west as well, but not exactly at the same location (panels D and I of Figure 1).

The presence of intense radiance near the patera's wall has been found to be common in many lava lakes (Mura et al., 2024b) and is suggested to indicate exposed lava due to the breakup of the crust near the patera's wall. A possible reason proposed for this phenomenon is a piston-like movement of the entire lake's surface, without lava spilling out of the patera's rim. This idea is supported by the observations made by the Voyager 1, Galileo and Cassini spacecraft, that the shape of the low-albedo region was unchanged between 1979 and 1997 (e.g., McEwen et al., 1998; Allen et al. 2013). However, given Loki Patera's size, which is more like a sea than a lake, it is also reasonable to consider an alternative hypothesis where the surface does not rise and fall uniformly. Instead, the phase of this movement could vary across the lake. Indeed, in orbit 57, the most intense edge is the south side, while in orbit 58, it is the north side close to the promontories. This variation may be partially due to the observation geometry, as the patera's walls can be very steep and high, potentially blocking part of the radiance from the lake sections closest to the walls (Mura et al., 2024b). It is also notable that the eastern edge, which should be visible without obstruction in orbit 58, does not show any significant excess of radiance as if there is no exposed lava close to the walls in that part of the lake. One possibility is that, instead of the piston movement, a sloshing-type movement occurs in Loki Patera. In this case, at a given time, some part of the crust might move vertically, while other sections remain relatively stable, thus not producing crust disruption and lava exposure. This model is not yet tested but may be viable, given that Io's intense tidal forces applied onto such a large lake could potentially drive such a phenomenon (tides at Loki Patera have been discussed by Matson et al., 2006; see their figure 14). Another possibility is that the temperature difference is due to the non-uniformity in the lava gas content and/or crust bulk density across the patera, as proposed by de Kleer et al. (2017). In a final scenario, this signature could reveal lava erupting from source regions either at the southwest or northwest, then extruding out and sweeping across the floor. Thermally, all these scenarios may appear similar, so distinguishing between them is challenging.

## 3.2 Modelling the crust cooling

We start the analysis by noting that the three zones introduced in section 2 (Figure 1) have temperatures that are systematically different: zone 1 is always colder, and zone 2 is always hotter. L-band and M-band derived temperatures are in very good agreement: the correlation coefficient is greater than 0.99, there is no systematic bias between the two (they have the same mean value). We conclude that the possible presence of small, cold islands among the lava lake (which is, indeed, true, as discussed later), or of hot cracks, or, finally, of some calibration residual of the instrument, is not affecting the estimation of the temperature in a significant way.

With this premise, we can say that the cooling of the crust, observed for 330 days, aligns with the temporal model proposed by de Kleer and de Pater (2017), which has been generalized and discussed in the Appendix A2. The model is represented by the three black dotted curves. The $t_0$ date of the model (i.e. the beginning of lava cooling) is not known, and it is found by best-fitting the model against the data. The three onset dates with respect to the last observation on April 9, 2024 ($t_0$) are: $t_{01} = t_0$ -481 days for zone 1, $t_{02} = t_0$ -373 days for zone 2, and $t_{03} = t_0$ -418 days for zone 3. In other words, at the time of the last observation, the crust was 481, 373 and 418 days old in the three zones, respectively. These results indicate that the temporal sequence of cooling initiation in the three zones is: zone 1 (south), then zone 3 (east), and finally zone 2 (north). This could be the effect of the propagation front of a resurfacing wave that moves in counterclockwise direction, the same as observed in the Galileo photopolarimeter/radiometer (PPR) data (Rathbun et al., 2002). The phenomenon will be further discussed in section 3.4, where we use the terms "resurfacing wave" and "phase velocity" to indicate the movement of the ideal instantaneous location where the surface of the lake starts to cool down, without implying any specific physical model; these will be introduced in the discussion (sections 4 and 5).

In the Appendix A2 we describe two possible scenarios for the cooling of the crust. Scenario A assumes the same initial conditions as in de Kleer and de Pater (2017) (i.e., fully molten magma); scenario B refers to a case in which 5 meters of magma are cooling on top of some pre-existing crust. While scenario A provides an adequate fit to the temperature data, scenario B does not. As summarized in Appendix Figure A3, the cooling in scenario B occurs too rapidly to match the observed temperatures. We therefore conclude that hypothesis B (lava cooling on top of preexisting crust) is not compatible with the observations unless the thickness of the lava is large; at least 5 meters of crust would have been produced for each cycle in that scenario.

## 3.3 Time variability of Loki Patera

The cyclic brightening of Loki Patera is thought to result from large-scale resurfacing processes within the lava lake; it is estimated that the entire cycle lasts approximately 540 days, aligning with observed brightening events before 2001 (Rathbun et al., 2002; Matson et al., 2006). De Pater et al. (2017) and de Kleer and Rathbun (2023) show that recent activity exhibits a slightly shorter periodicity (490 days) compared to what Rathbun et al. (2002) observed. Our observations do not cover a full cycle, but they are also consistent with this shorter period. In fact, the bright area is observed in the southwest as in previous observations; the hotspot observed in the western edge of Loki Patera in April 2024 is very close to the location where PPR observed the beginning of an eruption in late 1999. Furthermore, the observation of April 2024 was obtained approximately 490 days after our calculated previous eruption in December of 2022, suggesting that this spot is the beginning of a new eruption event.

There are different alternative explanations for the cyclic activity of Loki Patera, other than that it is driven by episodic lava flows (Matson et al., 2006). Gregg and Lopes (2008) proposed that a brief eruption would deposit minimal fresh lava on Loki Patera's floor; most new lava would form a shallow intrusion via channels and tubes, with a significant portion possibly draining back through the main vent, which corresponds to the thermally bright zone observed in the NIMS data. This model helps explain why lava has not overflowed the patera rim despite multiple brightening events, interpreted as eruption episodes. Rathbun et al. (2002) proposed that the brightening cycles are due to periodic crust overturns rather than lava flows resulting from cyclic eruptions. They also suggest that such periodic overturns can happen if specific conditions are met. The periodicity occurs because, initially, the solidified crust on the lake must be stable, but as it thickens and cools, it may become unstable. Sinking slabs of crust may thus initiate convection within a magma "sea" (Matson et al., 2006). This shift in stability might occur due to thermal contraction alone. In the case of typical terrestrial basalt, the solidified crust has a density significantly higher than that of the liquid beneath it, but the crust's porosity might contribute to its initial stability, preventing immediate overturn. De Kleer et al. (2017) modelled their observations according to this scenario and pointed out that the flooding one is not very plausible, because that would require two distinct pulses of magma reaching different parts of the patera at separate times to explain the ground-based observations by Keck. A foundering crust model makes it challenging to retain numerous islands that have remained unchanged for decades, unless the lava is excessively fluid (Howell et al., 2014), which is possible with ultramafic lavas, which have been proposed but not confirmed for Io (Keszthelyi et al. 1998). We note that recent observations of Loki Patera with the JunoCam visible camera reveal a widespread bright, specular reflection across the patera floor at low phase angle, consistent with a fresh and regionally uniform and flat lava surface, typical of lava lakes but perhaps possible with extensive flows (Ravine et al., 2024).

Table 1 presents the total power emitted by Loki Patera. The blackbody radiant emittance (in W m$^{-2}$) is derived from the brightness temperatures using Stefan-Boltzmann's law and integrated over the entire lake. This calculation can be performed using either the M-band or L-band brightness temperatures (columns 5 and 6). As shown, the two values are in close agreement (a non-uniform temperature across the crust, such as in the case of frequent cracks, would result in a substantial difference between the M- and the L-band temperatures: see the calculation in the Appendix A4 about the central island). The single-band power output (in both the M and L bands, columns 3 and 4) does not correlate well with the total power, as previously observed by Mura et al. (2024a).

Table 1: total power output of Loki Patera at different epochs as estimated from JIRAM data.

| Date | Orbit | M band power (GW) | L band power (GW) | M-band derived tot. power (TW) | L-band derived tot. power (TW) | |
|---|---|---|---|---|---|---|
| 2022-12-14 | 47 | * | ** | * | ** | * M-band data is saturating and sub-pixels, ** L-band data is not available |
| 2023-05-16 | 51 | 135 * | 15 | 12.8 | 13.5 | * M band is saturating |
| 2023-10-15 | 55 | 41 | 3 | 8 | 8.4 | |
| 2023-12-30 | 57 | 29 * | ** | 7.8 | ** | * M band covers only ~3/4 of Loki P. ** L band not available |
| 2024-02-03 | 58 | 29 | 2 | 7.1 | 7 | |
| 2024-04-09 | 60 | 26 * | 2 | 6.7 | 7.1 | Affected by a small but intense outburst, (* which is saturating in the M band). |

Once the temperature temporal profile is modelled, and the phase of the onset across the lake is determined from data, calculating the total heat output from Loki Patera becomes straightforward, as demonstrated by Matson et al. (2006). We replicated this specific theoretical calculation based on our data, and the result does not differ substantially from that of Matson et al. (2006). Moreover, we note that the empirical total power output values we observe (Table 1) are in very good agreement with the theoretical model. Additionally, it is possible to calculate the time-averaged total power, yielding a value of ~10 TW, which aligns very well with previous literature (e.g. de Pater et al., 2017, Table 9).

## 3.4 Propagation of the resurfacing wave

De Kleer and de Pater (2017) modelled the overturn process proposed by Rathbun et al. (2002), finding a propagation speed of 1 km per day. Such speed is highly sensitive to the properties of the magma or the physical process by which the crust recycles, which suggest it could differ between brightening events (Rathbun and Spencer, 2006). The varying resurfacing rates between the northern and southern regions suggest differences in magma properties, such as composition or gas content, rather than local topography, since the floor of the patera is relatively flat; the southern part of the patera has a resurfacing rate twice that observed in 2001, indicating more variable lava properties (Howell and Lopes, 2007). In our study, we do not observe such a difference, instead the northern part has a faster resurfacing rate as discussed in the Appendix A3.

In de Kleer et al. (2017), two overturn waves are suggested. The first one was moving in clockwise direction from the northwest at 1 km/day, while the second one was moving in counterclockwise direction from the west at 2 km/s and started 70 days later. The two fronts should meet approximately at the east or southeast, where our zone 3 is. However, the timings in which the resurfacing wave arrives to the three locations in this study indicates that only single wave, moving counterclockwise from the eastern part of the lake, was present (this is also obvious by looking at radiance maps, where the north, and not the east, is the brightest region). The distance from zone 1 to zone 3 is approximately 100 km, and it is 120 km from zone 3 to zone 2. Hence, the timings are consistent with a phase velocity of ~2 km/day between zone 1 and 3 and ~ 3 km/day from zone 3 to zone 2 (there are, in fact, ~65 days between $t_1$ and $t_3$ and ~45 days between $t_3$ and $t_2$). This implies that either the clockwise wave was not present in the event we observed, or it started much later than the counterclockwise one (and not earlier as reported by de Kleer et al., 2017). We conclude that each resurface event may have its own characteristic as proposed by Rathbun and Spencer (2006); this is also in agreement with de Pater et al. (2017), who reviewed 30 years of observations of Loki Patera, and noted a reversal in the direction of flow propagation. Since 2009, the flow has moved in a clockwise direction, beginning in the north or northeast corner, and spreading toward the southwest. In contrast, during the Galileo era, the flow propagated counterclockwise, starting in the southwest and moving toward the east. We are now once again observing a counterclockwise propagation.

To further analyze the propagation of the resurfacing wave, we present a similar calculation in the Appendix A3, but repeated across 15 different regions. The results are evident: the progression of the resurfacing wave is consistently revealed by the data, indicating an average speed of 2 to 3 km/day, with a slightly higher velocity observed in the northern region. The crust thickness at the time of the last observation ranges from 5 to 6 meters.

## 3.5 The central island

A persistent island in the middle of Loki Patera has been observed since the Voyager 1 flyby of March 1979, remaining stable despite surrounding lava eruptions. This island is likely an anchored feature prominent above the floor and the lava surface, rather than a floating raft, and similar islands are seen in other paterae such as Tupan (Lopes et al., 2004). The JIRAM data reveal no visible channels or cracks on the central island, except for one located in the south, which appears in the M-band radiance maps from both orbits 57 and 58. This feature resembles a thin arc extending from the northwest to the southeast. Aside from this, the island appears largely featureless. In contrast, the image in the visible spectrum displays several channels where thermal emission was previously detected in the infrared (Lopes-Gautier et al., 2000) and could potentially be detectable by JIRAM, assuming they still exist. We propose that one possible explanation for JIRAM's inability to detect these channels or cracks is their depth, combined with the misalignment of their axes relative to the line of sight during the observation. Estimating the depth of these features is challenging, as it requires knowledge of the channel width and the inclination of its walls — parameters that remain unknown. In the Appendix A4 we will show that, in fact, there is some extra emission from some of these channels, although very faint.

The combined analysis of the L- and M-band radiances in a fashion similar to that performed in Figure 2 (that is, defining a zone inside the island and calculating the temperature function of time) is quite challenging because the radiance is very close to the noise level. However, we note that, on average, the L-band derived temperatures are about 20% higher than the M-band ones (255 K vs 210 K); a possible explanation is that there are many small areas (sub-pixel) that contribute to the average radiance without being visible in the image (as detailed in the Appendix A4). This hypothesis is supported by the fact that the central island is significantly warmer than the region that surrounds Loki Patera as would be expected if the island was being heated by the active lava lake.

## 3.6 Minor islands

Numerous smaller colder features were observed by Voyager 1 in different locations across the lake, but it was uncertain whether they were actually anchored to the floor. The comparison with recent JIRAM data shows that at least 20 of them are most likely real anchored islands since they could not have remained in the same position if they were floating. Some minor evidence suggesting that these features are anchored to bedrock, rather than floating crust, is the fact that their temperatures match the coldness of the surrounding plains.

The close distance between islands (separation is ~10 km), combined with their small size, points to a lake level that has not changed since 1979 (and this favors the hypothesis of the periodic overturns of part of the crust over that of the periodic flows, because the latter would constantly increase the level, except if there was a complete drainback). The supposed vertical up-and-down movement, proposed for other lava lakes on Io (Mura et al., 2024b) and necessary to produce the crust breakup close to the patera's walls, may not be sufficient to submerge all the island completely. In fact, it seems that in the subsequent observation (orbit 58, panel D of Figure 1) the larger islands are still visible; the smaller ones may not be revealed due to the lower spatial resolution of this map.

A periodic lava flow model is inconsistent with JIRAM's observation of small islands in the middle of the lake, where the flow should have passed. The lava is estimated to grow at least 5 meters per cycle (de Kleer et al., 2017). If multiple layers are flooded and if cycles have occurred throughout all the time from Voyager's observation to JIRAM's, the total thickness over 45 years would have been at least 150 meters (Howell et al., 2014), which is not a negligible figure compared to the small size of the islands. However, it is worth noting that our analysis cannot determine the height of the islands, and whether any islands have disappeared or how many might have, as it focuses solely on counting similar features. In addition to the current analysis, it is pertinent to consider the characteristics of very small, closely packed islands. Are they anchored to the floor of the lava lake? We suggest that this is the case but, if so, why have the subsurface parts of each island not been thermally eroded so that they have collapsed during the multiple resurfacing cycles? Superficially, these islands exhibit similarities to those found in terrestrial lagoons, which are typically situated in shallow waters. Alternatively, they may be analogous to pillars ("inselbergs") formed in terrestrial karst topography near the coast (Veress, 2020). This observation raises a hypothesis regarding the temporal variability in the depth of Loki Patera's lake, but which has yet to be quantified.

## 4   Discussion

There are at least four experimental pieces of evidence from Juno/JIRAM observations that can be used to discuss previous models of Loki Patera:

i. The crust does not exhibit a continuous, hotter ring near the edges of the patera, as observed in other lava lakes on Io (Mura et al., 2024b). A more complex pattern is visible near the wall, analogous to localized features observed by Galileo. However, the radiance maps do not display the clear characteristics of lava lakes found elsewhere.
ii. Many small, closely spaced islands have persisted for at least 45 years.
iii. The phase velocity of the resurfacing wave is highly consistent, with only one wave observed, different from what was observed by de Kleer et al. (2017).
iv. The temporal profile of crust cooling is slow, suggesting that a significant amount of magma beneath the crust is needed to provide the latent heat necessary to decelerate such cooling.

In addition, data from April 9, 2024, might indicate the onset of a new thermal wave. However, this data could suggest a localized point source rather than widespread crustal movement or coverage, potentially related to a conduit feeding Loki Patera, if such a conduit exists. A small point source was observed on Dec. 30, 2023, but later disappeared.

Alternatively, if the radiance burst observed on April 9, 2024, represents the front of a resurfacing wave, then the smaller burst from the previous orbit on February 3, 2024, could correspond to the position of the wave front two months earlier, at the start of the cycle. Both M and L bands exhibit similar data patterns, supporting that this is a real physical feature. The two locations are approximately 50 km apart, suggesting a phase velocity of 1 km per day, which is in line with the overturn model's proposed limits. If so, then region 1, whose crust is estimated to be 480 days old at the time of the last observation, is about to experience another resurfacing. Hence, the duration of the cycle estimated by JIRAM would be just above 480 days.

Point i) seems to rule out that the piston-model lava lake proposed by Mura et al. (2024b) could work for Loki Patera, as the characteristic ring is visible only in limited regions, which may be explained by a different crustal erosion mechanism. A piston model would imply the draining and refilling of a significant amount of magma, while warmer edges have been observed by Galileo/NIMS.

Regarding point ii), it is uncertain whether the presence of islands imposes limits on the resurfacing models; it certainly imposes constraints on the lake's depth. Kauahikaua et al. (1998) propose an erosion rate of 10 cm/day for basaltic lava streams in tubes at Hawai'i, but this value may not reflect conditions at Loki Patera. If taken as a reference, it would suggest that the islands should have disappeared between Voyager 1 and JIRAM observations. Of course, it is also reasonable to expect that the small islands did not form immediately before the Voyager 1 observations and so their longevity may be at least a century or more.

We can confidently state that most of the time, islands are surrounded by a layer of crust above the magma, and that the crust itself remains stationary. Erosion likely occurs only during the passage of the resurfacing wave, which is a minor portion of the time. However, the base of each island might well be subjected to thermal erosion by the hot magma beneath the crust. Nevertheless, many islands remain visible after 45 years. In addition to that, our analysis does not fully account for potentially disappeared islands, as we focus on stable features rather than those that have vanished, which are more challenging to track. On the other hand, in a continuously large-scale convective magma lake scenario, erosion would occur constantly, making island survival unlikely. Therefore, the presence of islands argues against the presence of a global sub-crustal convection, and in favor of a shallow lake. However, these islands can only be integrated into a model of Loki Patera after their formation mechanism is understood. The islands, including the large central one, could represent a portion of the surrounding terrain that became isolated during the subsequent formation of the lava lake. Alternatively, a possible mechanism by which the islands originated could be that they are remnants of crust that did not subside. In the context of a large-scale and catastrophic overturn of the crust, it is plausible to imagine that a few fragments of crust could remain buoyant. During the subsequent cooling and thickening of the crust, these remnants would systematically acquire a greater thickness compared to the surrounding areas. Over time, they could persist as crustal "icebergs," potentially reaching the seabed. One possibility is that the floor has varying elevations, so that the island could become attached to the shallowest portion. A significantly more comprehensive and complex modeling effort than that presented in this work is required, but we suggest that this hypothesis be further developed in the future, as it could explain, for instance, why all minor islands have similar dimensions.

Point iii) implies that whatever process drives the resurfacing wave, it affects the entire lake and propagates at an almost constant speed (2 km/day in the south, 3 km/day in the north). In the foundering crust scenario, the resurfacing rate reflects the bulk density of the crust. Independent magma sources along the western edge of the lake, with higher density in the southern source, account for the faster resurfacing previously observed there (Rathbun et al., 2002; de Kleer et al., 2017). Those authors proposed that the difference may be due to varying gas content in the magma, which affects the porosity and cooling rate of the crust. Higher bulk density causes the crust to sink faster after solidification. On the other hand, models suggesting magma outflow from a specific point (as might be observed in the April 9, 2024, map showing a nearly point-like source) must account for the magma covering 300 km in approximately 110 days. If the magma is less than 10 meters thick, it would solidify into crust within this 110-days period. The resulting crust thickness might raise the crust level incompatibly with the presence of islands.

Point iv) rules out the possibility of observing a thin layer of lava cooling over pre-existing crust. At least 10 meters of molten magma (and its crystallization latent heat) is required to explain the data, though it could be thicker. A different consideration leads to similar conclusions. Loki Patera has existed for at least 45 years; over long periods (decades), it should be in thermal equilibrium, meaning the heat coming from below the surface (from Io's interior) matches the heat being emitted from the surface. This balance involves an average of ~10 TW of power according to our and previous models (Matson et al., 2006; de Kleer and de Pater, 2017). However, over shorter time scales, there's a disequilibrium. After an eruption or resurfacing event, heat output is mostly during the initial phase, while the heat input from Io's interior does not follow this temporal profile, and supposedly remains relatively stable during this period. As a result, Loki Patera undergoes a cycle where for the first ~130 days after resurfacing, the surface emits more heat than the average input; for the following ~400 days, the heat emission decreases below the average. The magma beneath the surface hence should accumulate and release heat like a buffer. The energy involved over a full cycle is about $5\times10^{15}$ J/km². We do not know the depth of the lake but it's easy to calculate that a depth of less than 10 meters would result in a magma temperature variation of more than 200 K. Such a large temperature variation in the magma would challenge the validity of the model, which otherwise agrees well with the data (for instance, the 10 TW heat output matches observations very well). Although this criterion is not highly restrictive, we emphasize that with more precise measurements of lava temperature in the future, it may be possible to derive a more accurate estimate of the lake's depth.

In summary, point iv) suggests a minimum lake depth of a few 10s of meters, while point ii) indicates a physical limit to the lake's depth due to the presence of packed islands, though quantifying this limit is challenging.

# 5  Summary and conclusions

Loki Patera, one of the most prominent volcanic features on Io, stands out due to its unique thermal characteristics. Unlike smaller Ionian lava lakes, which often display continuous bright rings of intense heat, Loki Patera lacks this feature, although there is evidence of isolated brightenings from the lake's perimeter, both from JIRAM and in previous Galileo data. This absence might be explained by the sheer size of the lava lake, which could prevent its surface from moving as a single, cohesive unit, as seen in smaller lakes where the surface has been suggested to move more like a piston.

Another intriguing aspect of Loki Patera is the persistence of small islands within the lake. These islands, first observed by Voyager 1 roughly 45 years ago, remain in the same locations today. This constancy challenges different models discussed here: it would require extremely fluid lava for the surface flows to weave through them so evenly (Howell et al., 2014). It is also hard to understand, in the lava lake with a foundering crust model, how the wave of sinking crust could move smoothly around these obstacles (Howell et al., 2014). In fact, the models proposed by Rathbun et al. (2002), Matson et al. (2006), and de Kleer and de Pater (2017) did not account for the presence of islands. A refinement of these models, incorporating these stable features, is necessary to determine whether the crustal foundering scenario remains consistent with the observation of at least 20 (and likely many more) minor islands distributed throughout the lake.

The fact that the thermal waves, which move across the lake, seem unaffected by these islands suggests that they may be anchored into the lake's floor and do not disrupt the resurfacing process.

Additionally, the slow rate at which the crust of Loki Patera cools points to the presence of a hot, molten magma body beneath the surface that is at least a few tens of meters deep. This depth is necessary to provide the latent heat required to slow the cooling process. On the other hand, the long-standing existence of the small islands implies that thermal erosion at the base of these features is minimal, which fits with observations that their shapes have remained unchanged since Voyager's initial images. This is true for the large central island as well, which also shows little to no change over the same period.

In April 2024, two small new hotspots appeared west of the main island, which could indicate new vents breaking through the older crust. This suggests that Loki Patera's resurfacing process is ongoing and dynamic.

The fractures that crisscross the main island show little associated thermal activity. This raises the question of whether these fractures are too shallow to reach the lava beneath or if the infrared data from the JIRAM instrument might have barely detected thermal signatures due to viewing geometry.

Although there are no upcoming planned spacecraft missions to Io, one future objective could be to monitor the persistence of these small islands. Thermal erosion, in theory, should destroy such features relatively quickly, yet these islands have persisted for over four decades. Their continued existence would challenge current models of their structure, particularly below the lava lake's surface, as well as their mode of formation.

In the meantime, numerical modeling could provide valuable insights into the resurfacing process at Loki Patera. These models should consider the estimated depth of the lava lake, as well as the role these small islands play in disrupting the overturning or foundering of the crust. Additionally, the observations from April 2024 suggest that a new overturning cycle might be starting from a point source. Understanding how such events begin could help refine our understanding of the mechanisms driving Loki Patera's volcanic activity.

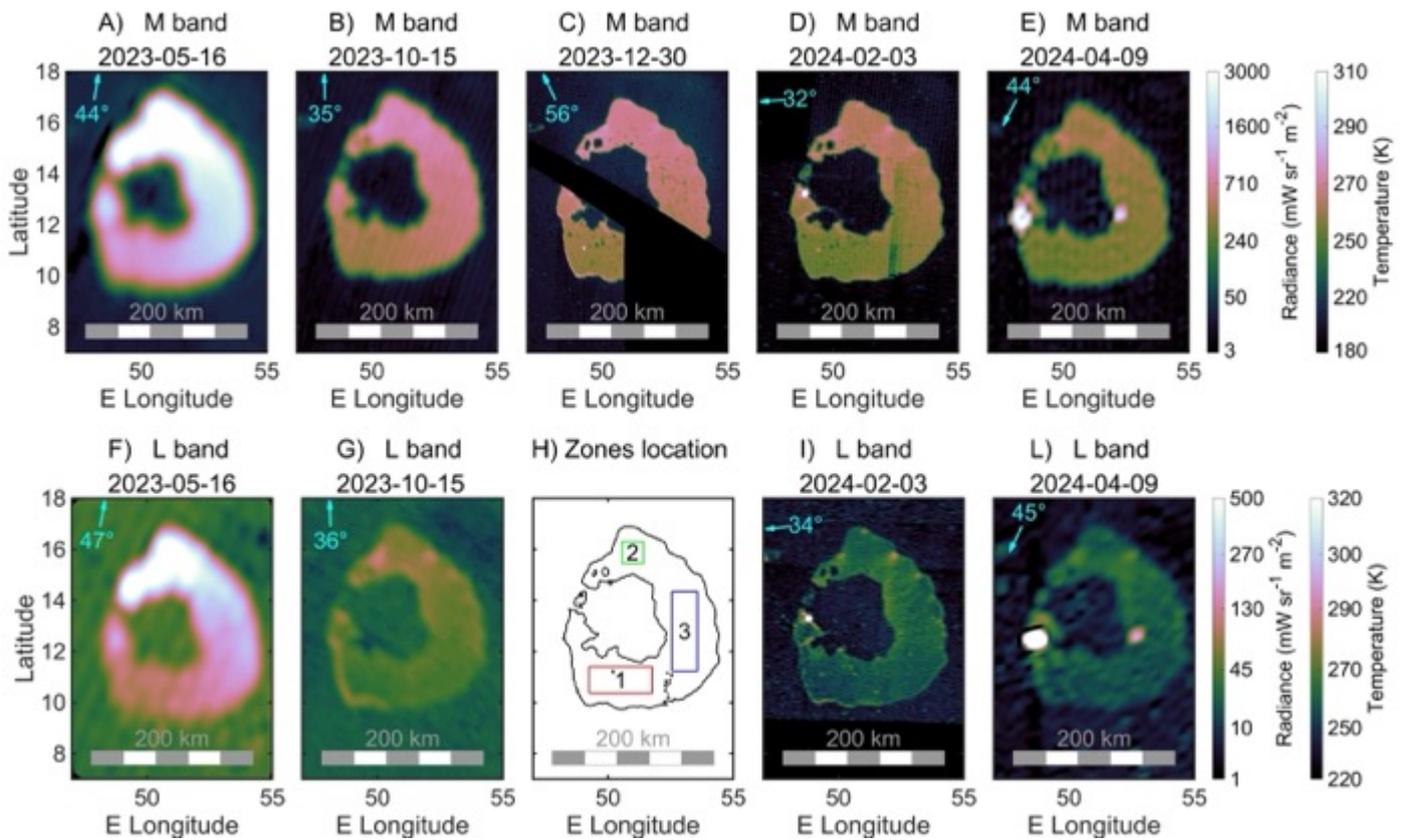

**Figure 1.** Top: M-band radiance maps of Loki Patera (integrated from 4.5 to 5 μm) taken during Juno's orbits 51, 55, 57, 58 and 60. The observations taken during orbit 47 are at very low spatial resolution (~100 km), and partially saturated, but it reveals the onset of the cycle. Observations from May 16 (orbit 51) are also partially saturated. Observations from orbit 58 (Feb. 3, 2024) have a stripe that is an instrument artifact. Observations from orbit 60 (April 9, 2024) are saturated in the bright feature at 49°E, 12°N in both bands. Bottom: L-band radiance maps (3.3 to 3.6 μm). Orbit 57 has no L-band observations; hence, we use that panel to show the location of zones 1, 2 and 3 (see text for details). The cyan arrows indicate the direction of the sub-spacecraft point at the time of the observation (that is, the approximate direction from which Loki P. is seen); the cyan numbers indicate the value of the emission angle (0° corresponds to looking at Loki P. vertically). Note that the color bars for the radiance are not linear; the supplementary color bars on the right show the corresponding brightness temperature associated with the measured radiances.

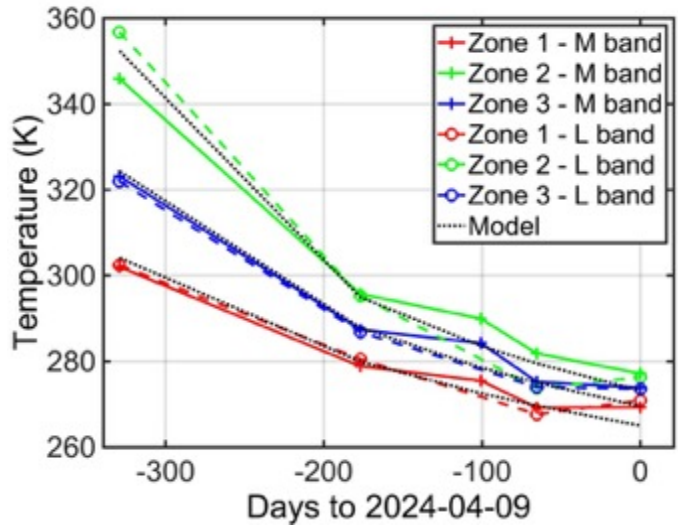

**Figure 2.** Temporal variability of the retrieved temperatures in the three zones. Data related to the first point of M-band in zone 2 is saturated; this explains why the M-band temperature (345 K) is lower than the L-band one (~355 K). The dotted lines represent the model fit in the three cases (see section 3).

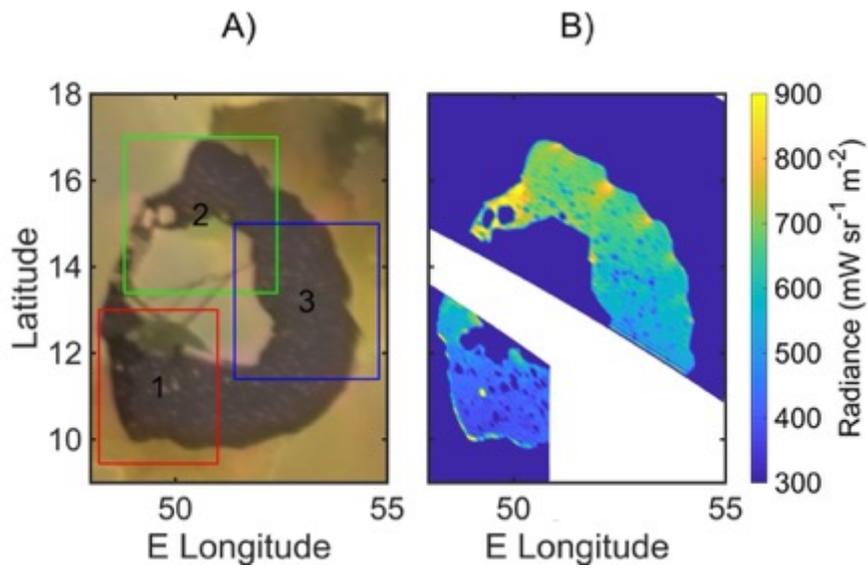

**Figure 3.** Loki Patera in the VIS (panel A) and in the M-band IR (panel B) as seen at high resolution in orbit 57. A small part of Loki P. (white region) was not imaged by JIRAM. Rectangles 1, 2 and 3 are indicated here and three magnified views of these regions will be shown in Figure 4. The colormap of the IR radiance map is tuned to enhance the details.

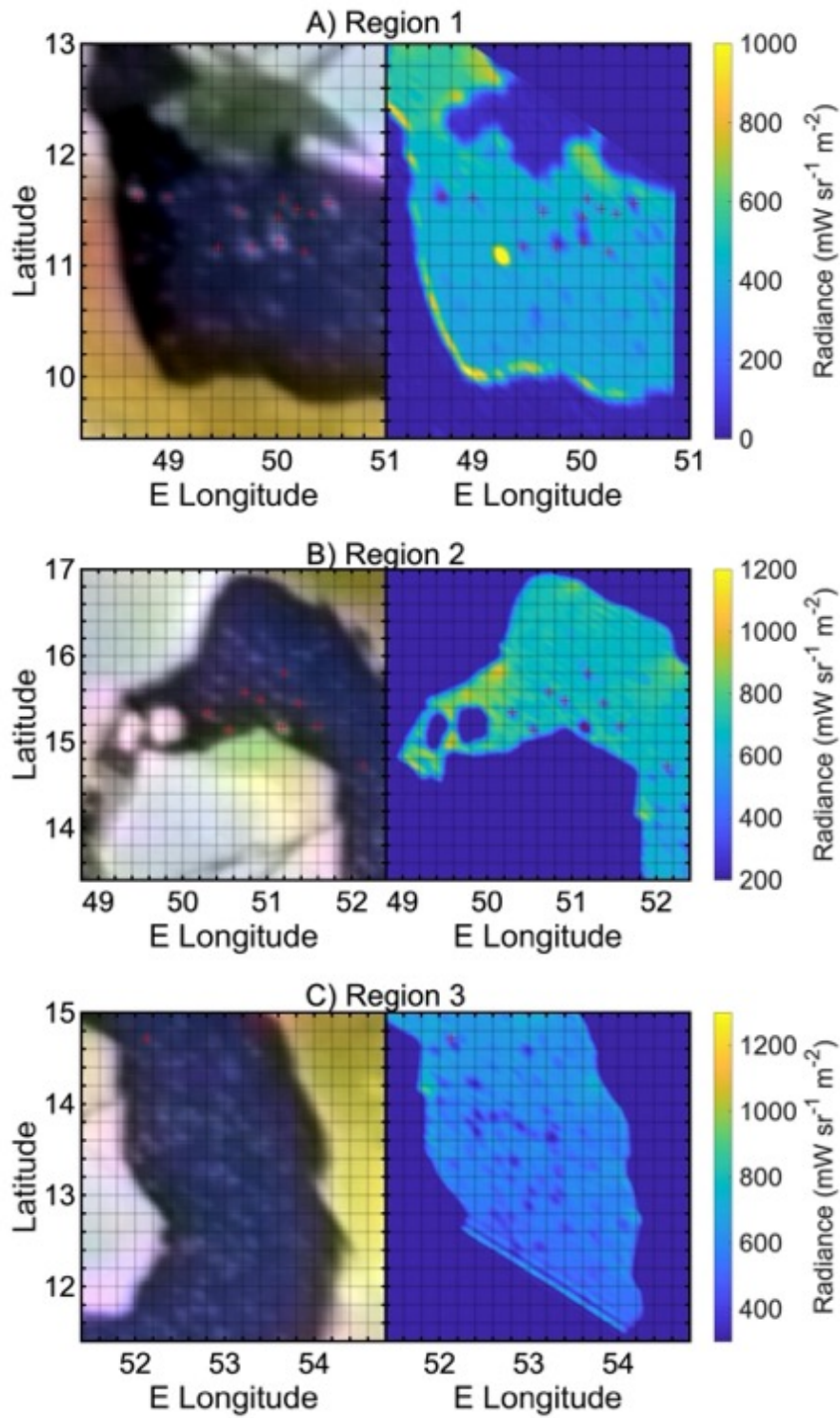

**Figure 4.** Visible and infrared maps of magnified regions 1 (panel A), 2 (panel B) and 3 (panel C). For each panel, the left part is the visible image, the right part is the M-band infrared radiance map. Colormaps of the M-band map differ from panel to panel to enhance the islands. Red dots indicate location of possible islands, identified in both the VIS and IR images, and are in the same exact places in both the VIS and IR maps

# Appendix

## A.1 Dataset

The imaging channel of JIRAM has two filters: the M band filter (~4.6 to ~5.1 µm) and the L-band (~3.3 µm to 3.6 µm). The spectrometer mode is not used in this study. The imager has an angular resolution of 237 µrad; JIRAM can take only one image and one spectrum for each rotation of Juno (~30 s). The FoV of JIRAM is 266 by 432 pixels, or 3.5° by 5.9°; the boresight is perpendicular to the spin axis of Juno. Hence, Io can be observed only when it crosses the spin plane of Juno (or when Juno is re-oriented to make it so). The de-spinning mirror of JIRAM was not functional at the time of the observations. We used very short integration times (2 ms on average) to keep the smearing low (0.8 pix/ms); residual smearing is corrected via software by applying a Lucy-Richardson algorithm (Lucy, 1974; implemented by MATLAB).

The NAIF-SPICE software (Acton, 1996) is used to map the images onto the surface of Io. For each orbit, we select the images with best quality and spatial resolution. Then, to increase the geometry reconstruction accuracy, the coalignment of these images is checked and, in case, corrected. Finally, to reduce the noise, co-aligned images are stacked (obtaining a 3D cube) and the median value along the 3$^{rd}$ dimension is taken, so to obtain one 2D super-image for each orbit and each band (as in Mura et al., 2020, 2024a, 2024a). After this procedure, the uncertainty about the geometry reconstruction of JIRAM images is estimated to be equal to, or less than, one pixel.

Table A.1: list of the observations used in this study.

| Juno's orbit | Date | Distance (km) | Resolution (km) | Number of M-band images used | Number of L-band images used |
|---|---|---|---|---|---|
| 47 | 2022-12-14 | 67'000 | 15 | 5 | 0 |
| 51 | 2023-05-16 | 51'000 | 11 | 6 | 2 |
| 55 | 2023-10-15 | 20'000 | 4 | 7 | 15 |
| 57 | 2023-12-30 | 3'500 | 0.4 | 2 | 0 |
| 58 | 2024-02-03 | 4'300 | 0.6 | 3 | 2 |
| 60 | 2024-04-09 | 23'000 | 5 | 1 | 1 |

In Table A.1 we report the date of observation, minimum distance (rounded), the spatial resolution at the surface (not considering possible smearing and the point spread function of the instrument, so technically that is a spatial sampling), and the number of images used for each band. The spatial resolution is calculated as iFoV × distance, that is the best resolution virtually obtained from that distance, and hence it is an indicative value.

## A.2 Model of magma and crust cooling

To analyze the dataset provided by the temperature as a function of time, we introduce a model that simulates the cooling and crystallization of magma. As the magma cools from the surface, it begins to solidify, forming a crust, while the deeper layers remain molten. The 1-D model employs the finite element method to solve the heat diffusion equation in the crust, following the approaches detailed by Lewis et al. (1996) and Strang and Fix (2008). The physical assumptions and constants in our model closely align with those described by de Kleer and de Pater (2017). In both models, the magma and crust density ($\rho$) are assumed constant at 2800 kg m$^{-3}$, with a latent heat of crystallization ($H_L$) set at 4×10$^5$ J kg-1. The thermal conductivity ($k$) is 0.9 W m$^{-1}$·K$^{-1}$, and the specific heat capacity ($c_p$) is 1500 J kg$^{-1}$·K$^{-1}$, ensuring that the fundamental physics governing the cooling process in both models remains essentially identical. A minor difference is that, while our model assumes constant magma and crust densities with temperature, De Kleer and de Pater (2017) consider temperature-dependent density. However, they also note that density variations have a minimal effect on the cooling dynamics and overall system behavior. Therefore, this approximation is not expected to significantly impact the model's accuracy or its predictions regarding cooling and crust formation.

At the surface, the Neumann-type boundary condition is governed by the Stefan-Boltzmann law, which describes radiative heat loss as a function of temperature. The heat flux is:

$$\phi_q = \varepsilon \, \sigma \, T^4,$$

where $\varepsilon$ is the emissivity of solid crust (0.9), $\sigma$ is the Stefan-Boltzmann constant, and $T$ is the temperature at the surface. The numerical solver uses a Galerkin method to iterate along the domain of time; the time step is chosen to maintain the required stability with the criterion that the dimensionless parameter $\beta$ is less than 0.5:

$$\beta = \frac{k}{\rho \, c_p} \frac{dt}{dx^2} < 0.5;$$

the model simulates a 10-meter-deep column, with a 10-cm node spatial step ($dx$).

In the first scenario, the entire column consists of hot magma cooling only from the surface; the domain is initially set at a uniform temperature of 1475°C. The release of latent heat during crystallization slows the cooling process as the top layers solidify. The simulation lasts for 1000 days, which is longer than the proposed period for Loki's activity (~500 days).

In summary, the first scenario follows de Kleer and de Pater (2017) and produces the same results; only the numerical solver is different. However, an alternative scenario is modeled with altered initial conditions, where the top 5 meters consist of freshly deposited magma at 1475°C, while the bottom 5 meters represent pre-existing solid crust at 270 K (a conservative value: the lowest measured temperature is less than 270 K, and at the time of our last observation the crust was still cooling). This setup simulates the case of magma flooding over a cooler crust. As in the first scenario, the latent heat of crystallization is included, slowing the cooling rate during the solidification process.

The results of these two scenarios are presented in Appendix Figures A1 (temperature as a function of depth and time, and crust thickness) and A2 (surface temperature as a function of time). Scenario A remains virtually out of equilibrium even after 1000 days, with molten magma still present at the bottom of the simulation domain (this suggests that simulating depths greater than 10 meters is unnecessary).

For scenario B, we observe that after $t_0+100$ days, the top 5 meters of magma have solidified, and the latent heat has been fully released. From this point onward, cooling proceeds much more rapidly, as the absence of latent heat allows for faster temperature decreases. Initially, the two temperature profiles remain similar, but they diverge significantly after 100 more days ($t_0+200$ days). At this point, the cooler crust below continues to act as a heat sink, accelerating the cooling of the system. After a further 200 days ($t_0+400$ days), the surface temperatures in the two cases differ by 200 K or more, allowing us to compare the model predictions with observational data to assess which scenario is in better agreement with the temperatures measured by JIRAM.

The results of fitting both scenarios to the data are presented in Figure 3. The left panel (A) shows the same result as Figure 2 in the main body of the article. The only difference here is that, instead of displaying absolute time on the *x*-axis and showing the three models, we plotted relative time (showing only one model) and adjusted the data by subtracting the $t_0$ values derived from the best fit. The right panel (B) presents scenario B with adjusted $t_0$ values, though the results remain inconsistent with the data. The issue lies in the fact that the data exhibit a cooling rate of only 0.1 K/day in the region below 300 K, while the model predicts approximately double that rate.

Incidentally, we also tested a third case in which the starting condition is a hot column of solid crust. Regardless of the initial temperature, this simulation produces results that are entirely inconsistent with the data. This simulation does not have a clear physical interpretation, as it does not adhere to any particular model for Loki Patera; it was performed solely to demonstrate that latent heat is the only factor capable of explaining the temporal evolution of temperature.

## A.3 Detailed analysis of crust thickness function of location

We extend the analysis presented in the main part of this paper and illustrated in Figure 2 by dividing the lake into 15 distinct regions instead of the original 3. These regions are numbered sequentially in a counterclockwise direction, beginning from the southwest (right panel of Appendix Figure A4). Region 3 corresponds to the southern part of the lake, region 9 to the eastern part, and region 15 to the northern part. The regions are spaced with margins from the patera's wall to exclude extra radiance originating from near the walls. For each of these regions, four or five mean temperatures in both bands were derived from radiance maps across four or five different epochs/orbits (during orbit 57 regions 4 to 7 were not observed in the M-band, and the whole lake was not observed in the L-band). For each epoch and region, the average between the L-band and M-band derived temperatures was calculated. Using the method detailed in the main text, we estimate the cooling onset time, or crust age, by fitting the temperature data to a time-dependent model. The estimated crust ages are shown in panel A of Appendix Figure A4, and panel B illustrates the locations of the 15 regions. The regions are of approximately equal size (~1000 km$^2$) and spacing (~20 km). The temporal progression in panel A of Appendix Figure A4 suggests a faster cooling wave beyond region 9 (east). The total distance covered by the cooling wave between regions 1 and 15 is ~300 km over ~110 days, resulting in an average phase velocity of slightly less than 3 km/day.

The uncertainty associated to the estimation of the crust ages is not easy to assess, because several factors may play a role. We assume that the uncertainty on the input data (i.e. the temperature) is small because, as mentioned, L-band and M-band derived temperature align well. The uncertainty arising to the best-fitting method can be estimated by running a simple MCMC (Markov-Chain Monte Carlo) instead of performing a single best-fit of the crust age.

The standard deviation of the output is about 20 days (it changes slightly from zone to zone), and it is used as a reference value for the error bars in Appendix Figure A4.

## A.4 Central island

Appendix Figure A5 presents the radiance map of the central island in Loki Patera, captured in the M band. This map represents the same data as in Figure 1 but with a different zoom and color scale. In visible light, the central island appears to be traversed by several cracks or channels. However, these features are barely discernible in the infrared, despite the resolution seemingly being sufficient to detect them. As illustrated in Figure 1, Juno's viewpoint is considerably inclined. In orbit 57, the emission angle is 56°, with a view from the northwest of the lava lake. In orbit 58, the angle decreases to 32°, and Loki is observed from the west.

In neither case are the channels strongly emitting in the infrared, except for an arc in the southwest, which appears to correspond to a visible feature but does not match exactly. This arc is clearly visible in orbit 57 and can also be detected, with slightly lower radiance, in orbit 58. Since the viewing angle in orbit 57 is more aligned with the direction of this supposed channel, we infer that it is deep. The more aligned the line of sight is with the channel, the less shadow the walls cast on the warmer, lower channel floor compared to the rest of the island.

A horizontal feature is observed in orbit 58, though it may be a JIRAM artifact, as it is aligned with one of the sensor's axes. A barely noticeable diagonal feature, running from southwest to northeast and more visible in the raw data (Appendix Figure A6), aligns remarkably well with a feature observed 45 years ago in Voyager 1 images, which was thought to be one of the wider channels. In summary, there is sufficient indication that the bottom of these channels is emitting at higher temperature. We also conclude that the channels visible in the optical spectrum are likely deep, such that they can only be discerned in the infrared when viewed at a low emission angle, or with a line of sight that is parallel to the direction of the channel.

The island appears hotter in orbit 57 than in orbit 58, suggesting the possibility for a temperature-versus-time analysis for this region similar to that in Figure 2. However, the low average radiance makes the data quite noisy, which could result in a misleading analysis. Also, at very low temperature the subtraction of reflected sunlight is more challenging and may leave some residual contribution.

Notably, temperatures derived from the L-band are consistently higher than those from the M-band by about 20%. A possible explanation for this discrepancy — aside from the obvious hypothesis of calibration residuals, which however have been ruled out by previous analyses in this study — could be non-uniform temperature distribution across the island. This may result from the presence of sub-pixel hot spots. For instance, if most of the island is at 200 K, with just 0.01% consisting of hotter features at 600 K, the overall radiance would correspond to that of a blackbody at 215 K in the M-band and 255 K in the L-band, thus explaining the 20% difference. Although the actual scenario is likely more complex, with many possible physical configurations, this example illustrates the reasoning behind such a hypothesis.

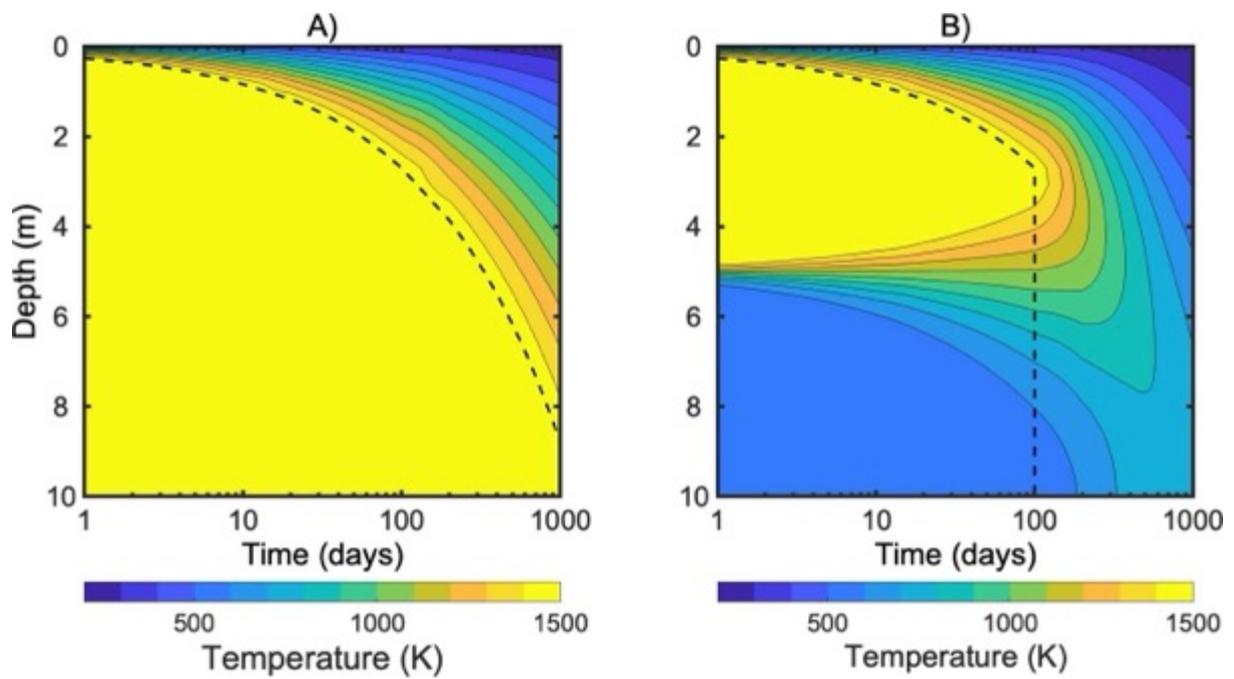

**Figure A1.** Temperature of a 10-meter column of cooling magma as a function of depth and time. Scenario A begins with a uniform 10-meter column at 1475 K; Scenario B begins with a uniform 5-meter column at 1475 K. The dashed line represents the depth of the crust at any given time.

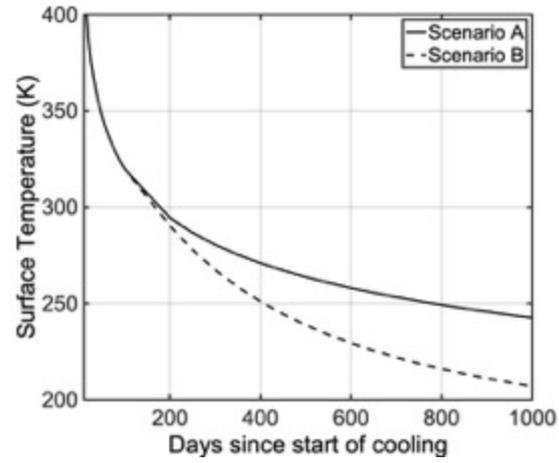

**Figure A2.** Surface temperature versus time for the two different scenarios. Solid line: scenario A; dashed line: scenario B. For scenario B, after ~100 days since the start of the surface cooling, the latent heat of crystallization is exhausted, and the surface temperature begins to drop faster than in Scenario A. This becomes more evident after ~200 days.

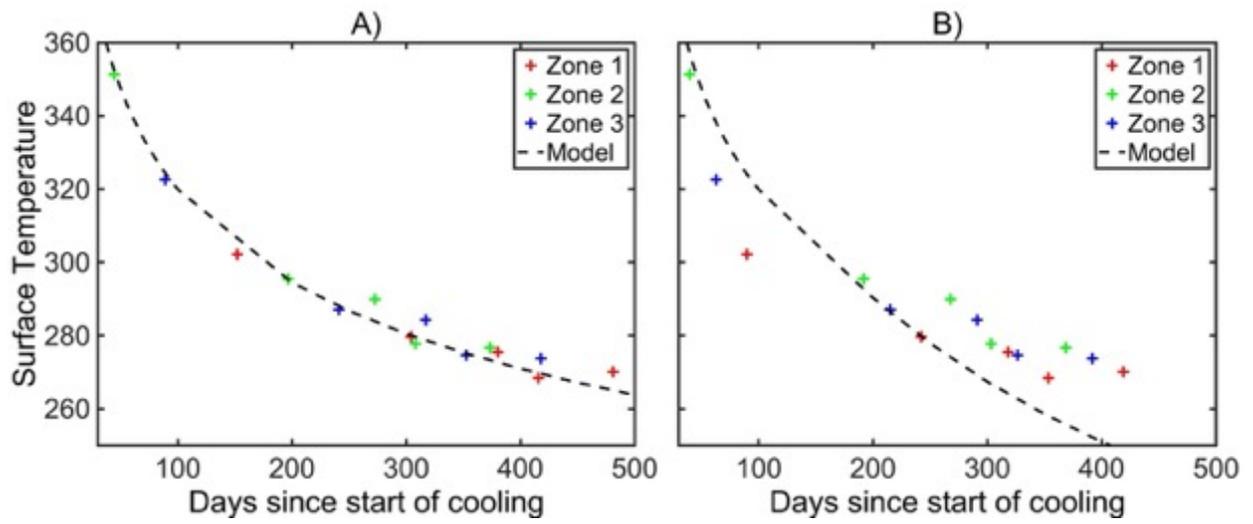

**Figure A3.** Panel A: comparison between data and model, with initial condition as in scenario A. Panel B: same for scenario B. Crosses represent the 15 data points (3 zones × 5 epochs), and colors are chosen according to Figure 2.

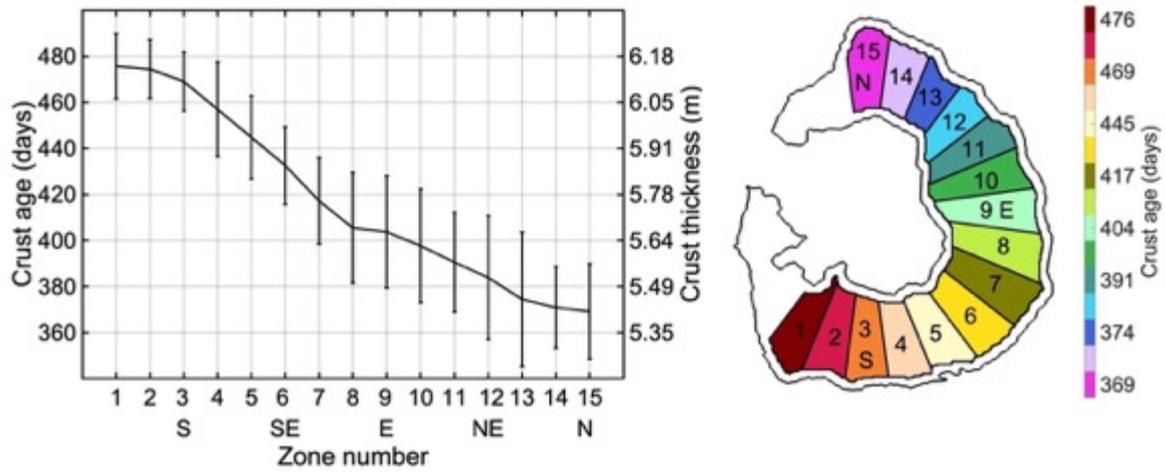

**Figure A4.** Panel A: crust age and thickness for 15 different zones of Loki Patera. Panel B: location of the zones chosen for the analysis. The color of the zones is coded according to the age of the crust, as indicated by the color bar to the right. The reference time for the crust age is set to the final observation epoch: April 9, 2024.

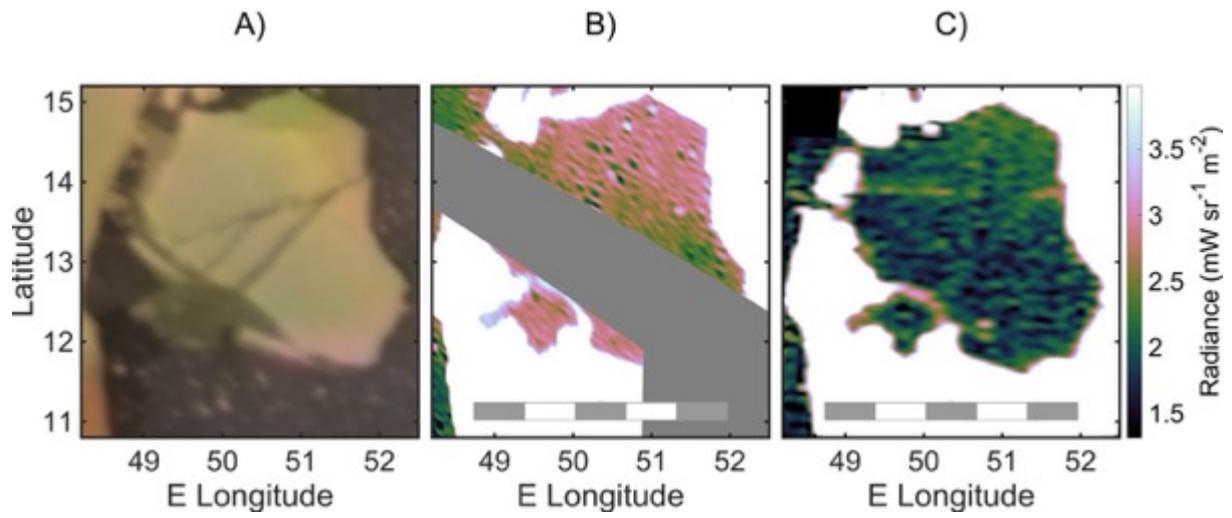

**Figure A5.** Visible image of Loki Patera (panel A), and M-band radiance maps of the central island a (integrated from 4.5 to 5 μm) taken during Juno's orbits 57 and 58 (panels B and C).

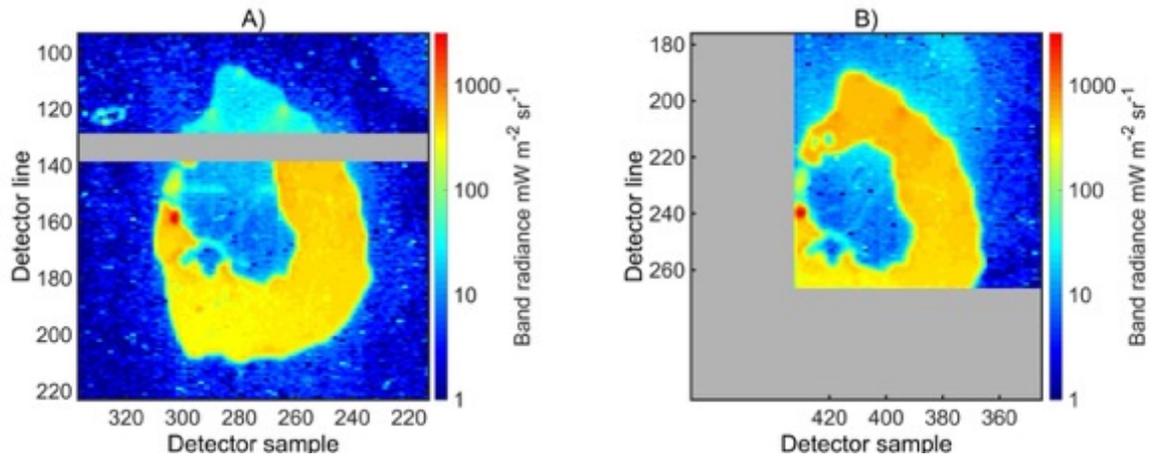

**Figure A6.** Two unprocessed and unprojected images of Loki P. taken during orbit 58 (2024/02/03; panel A at 17:54:10 and panel B at 17:54:40). Only a square 2×1 pixel filter has been applied to remove the odd-even effect. In panel A, the gray band is a 10-pixel wide unactive region of the detector, which separates the part of the detector that is placed beyond the L-band filter (lines from 1 to 128) and beyond the M-band filter (lines from 139 to 266). In panel B, gray area is outside the detector. A faint, diagonal line of brighter radiance, going from the southwest part of the central island to the northeast one, is visible in both images. The horizontal region in the central island, at line ~150, is probably an artifact of the detector since it is visible only in panel A. A second, fainter diagonal line, going from west-southwest to northeast, is barely visible in panel B.

# 6 Open Research

The JIRAM dataset used for our analysis is publicly available at the Juno Archive at the Planetary Atmospheres Node (Data from orbit 60 will be available on Oct. 20, 2024):

https://pds-atmospheres.nmsu.edu/PDS/data/PDS4/juno_jiram_bundle/data_calibrated/.

The Voyager-Gaileo data used for our analysis is available through Williams et al., 2011a, 2011b.

# 7 Acknowledgments

We thank Agenzia Spaziale Italiana (ASI) for the support of the JIRAM contribution to the Juno mission. This work is funded by the ASI–INAF Addendum n. 2016-23-H.3-2023 to grant 2016-23-H.0. Part of this work was performed at the Jet Propulsion Laboratory, California Institute of Technology, under contract with NASA.